\documentclass[aps,prl,superscriptaddress,showkeys,showpacs,twocolumn,longbibliography,nofootinbib]{revtex4-1}
\usepackage[utf8]{inputenc}
\usepackage{amsmath}
\usepackage{amsfonts}
\usepackage{amsthm}
\usepackage{amssymb}
\usepackage{float}
\usepackage{braket}
\usepackage{graphicx}
\usepackage{url}
\usepackage{amscd}
\usepackage{physics}
\usepackage[colorlinks=true, pdfstartview=FitV, linkcolor=red, citecolor=blue, urlcolor=blue]{hyperref}
\usepackage{slashed}
\usepackage[normalem]{ulem}

\newcommand{\Ldyn}[1]{L_{\mathrm{dyn},#1}}
\newcommand{\Lext}[1]{L_{\mathrm{ext},#1}}
\newcommand{\jext}{j_{\mathrm{ext}}}

\graphicspath{{./figures/}}
\begin{document}
\title{Quantum-classical simulation of quantum field theory by quantum circuit learning}

\author{Kazuki Ikeda}
\email[]{kazuki.ikeda@stonybrook.edu}
\affiliation{Center for Nuclear Theory, Department of Physics and Astronomy, Stony Brook University, Stony Brook, New York 11794-3800, USA}
\affiliation{Co-design Center for Quantum Advantage, Department of Physics and Astronomy, Stony Brook University, Stony Brook, New York 11794-3800, USA}

\begin{abstract}
We employ quantum circuit learning to simulate quantum field theories (QFTs). 
Typically, when simulating QFTs with quantum computers, we encounter significant challenges due to the technical limitations of quantum devices when implementing the Hamiltonian using Pauli spin matrices. To address this challenge, we leverage quantum circuit learning, employing a compact configuration of qubits and low-depth quantum circuits to predict real-time dynamics in quantum field theories. The key advantage of this approach is that a single-qubit measurement can accurately forecast various physical parameters, including fully-connected operators. To demonstrate the effectiveness of our method, we use it to predict quench dynamics, chiral dynamics and jet production in a 1+1-dimensional model of quantum electrodynamics. We find that our predictions closely align with the results of rigorous classical calculations, exhibiting a high degree of accuracy. This hybrid quantum-classical approach illustrates the feasibility of efficiently simulating large-scale QFTs on cutting-edge quantum devices.
\end{abstract}

\maketitle
\emph{Introduction.}---
The exploration of quantum field theories (QFTs) has long been a cornerstone of theoretical physics, enabling us to delve into the fundamental interactions that govern the behavior of matter and energy at the smallest scales. While the mathematical framework of QFTs has proven to be exceptionally powerful in describing these phenomena, simulating them on classical computers has often posed formidable challenges, limiting our ability to explore and understand the intricate dynamics of quantum systems.

In recent years, the advent of quantum computing has promised to revolutionize our approach to simulating QFTs~\cite{Klco:2018kyo, Butt:2019uul, Magnifico:2019kyj, Shaw:2020udc, Kharzeev:2020kgc, Ikeda:2020agk, Rigobello:2021fxw,Ikeda:2023zil,Bauer:2022hpo,10.21468/SciPostPhys.14.5.129,Farrell:2023fgd,Bauer:2023qgm}. However, the technical limitations of quantum devices, particularly when implementing the Hamiltonian using Pauli spin matrices, have hindered progress in this field. This paper introduces a novel approach that harnesses the potential of quantum circuit learning (QCL)~\cite{PhysRevA.98.032309,havlivcek2019supervised} to address these challenges and offer a more efficient and accurate method for simulating QFTs. 

Our method leverages a compact qubit configuration and low-depth quantum circuits, allowing us to predict real-time dynamics in QFTs that requires a large number of qubits and high-depth of quantum circuits. One of the standout advantages of this approach is its capacity to make accurate predictions for various physical parameters using just a single-qubit measurement. To demonstrate its effectiveness, we apply this technique to predict quench dynamics, chiral dynamics, and jet production in a 1+1-dimensional model of quantum electrodynamics developed in~\cite{Kharzeev:2020kgc,PhysRevLett.131.021902}, where simulations of chiral magnetic effect and the quark-antiquark production in $e^+e^-$ annihilation proposed were performed based on Quantum Chromodynamics (QCD)~\cite{Fukushima:2008xe,PhysRevLett.107.052303,Kharzeev:2015znc,Kharzeev:2013ffa,Casher:1974vf,Loshaj:2011jx,Kharzeev:2012re}. 

Remarkably, our predictions closely align with results obtained through rigorous classical calculations, underscoring the high degree of accuracy that can be achieved.

This hybrid quantum-classical approach not only represents a significant advancement in the simulation of QFTs but also underscores the feasibility of efficiently simulating large-scale quantum field theories using cutting-edge quantum devices. By combining the strengths of quantum computing with classical techniques, our research opens up new horizons in the study of quantum systems and provides a promising avenue for the exploration of complex physical phenomena. Our approach will also leverage the use of machine learning applications in the field of condensed matter, nuclear physics, high energy physics, chemistry, biology and information science, enhancing our ability to analyze and interpret data~\cite{doi:10.7566/JPSJ.86.063001,PhysRevD.98.046019,KHAKZAD2023925,Boehnlein:2021eym,Felser:2020mka,2023PhLA..48829138N,2023arXiv230209751N,2023arXiv230617214N,https://doi.org/10.1002/qute.201900070,PhysRevX.8.021050,2019arXiv190912264V,2019SciA....5.2761H,PhysRevA.98.012324,PhysRevLett.121.040502}.

The main contributions of our work to quantum simulations of QFTs are summarized as follows:
\begin{enumerate}
    \item Using a three or five-qubit QCL, we were able to efficiently predict the real-time dynamics of 1+1d QED (up to 18 qubits), including quench, chiral dynamics, and jet generation. 
    \item The real-time evolution of several physical observables of total coupling, including energy and electric field, was predicted with good accuracy by reading out only one qubit.
\end{enumerate}
Our work will present a benchmark showing that the complex dynamics of a generic large quantum many-body system can be efficiently predicted by QCL with a small number of qubits and low-depth of circuits.

\begin{figure*}
    \centering
    \includegraphics[width=0.49\linewidth]{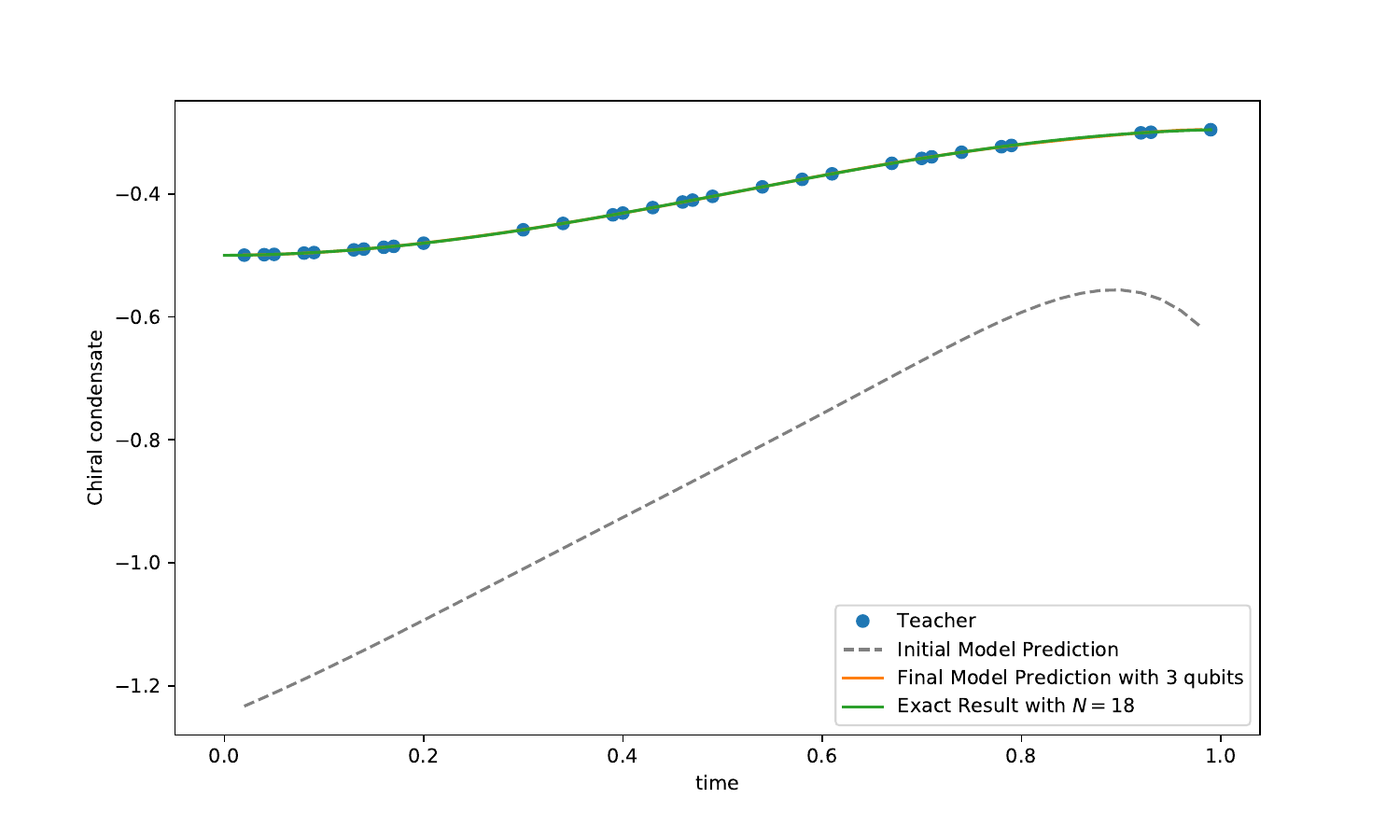}
     \centering
    \includegraphics[width=0.49\linewidth]{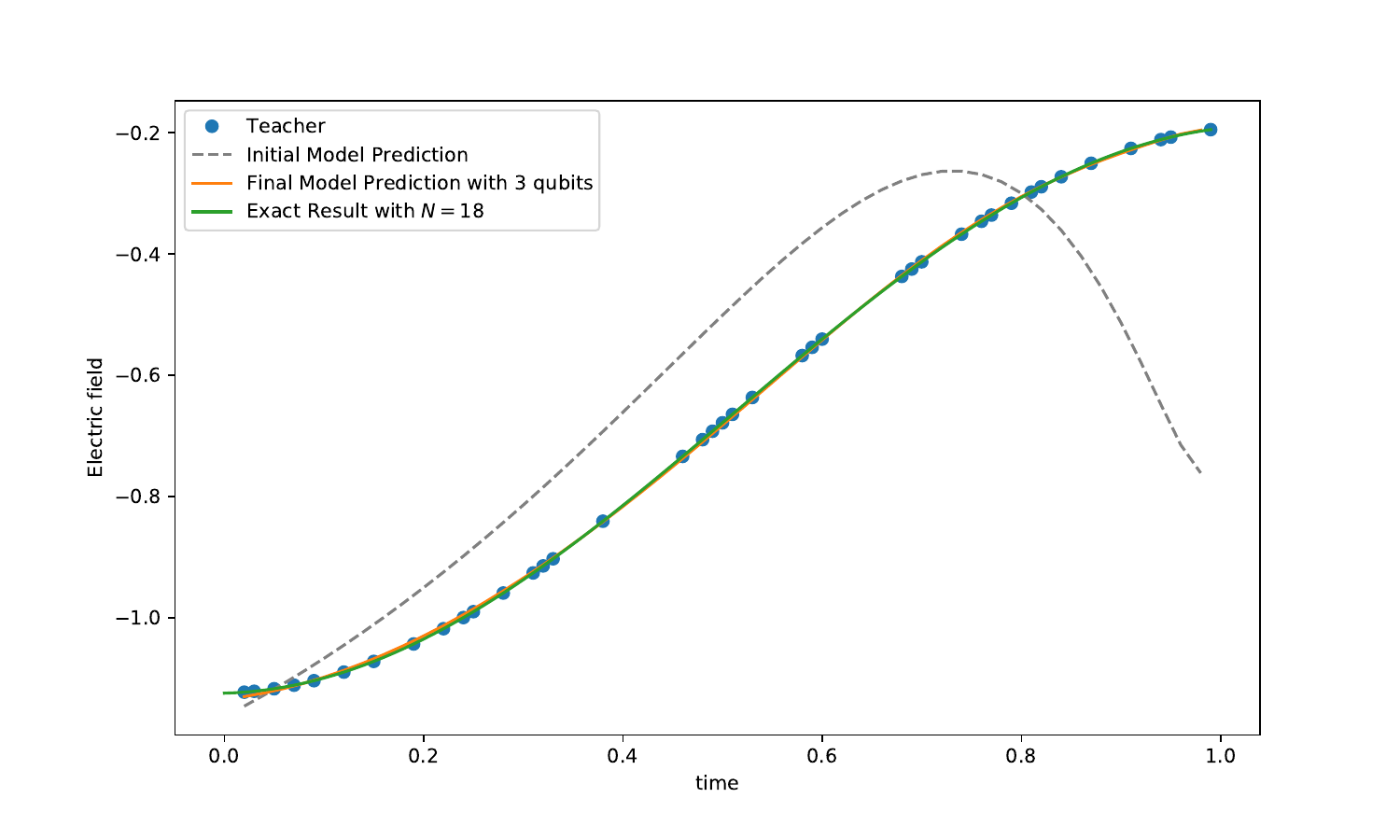}
    \caption{Predictions of the chiral condensate [left] and the electric field [right].}
    \label{fig:Neel}
\end{figure*}
\emph{Simulating QFT by QCL.}---
In quantum circuit learning, we follow a step-by-step process to train a quantum model. Here's a breakdown of the key components. We begin by preparing a set of training data, denoted as $\{ (x^{(i)}, y^{(i)}) \}$. Here, $x^{(i)}$ represents the input data (teacher data), and $y^{(i)}$ is the correct output data we expect the model to predict.

Next, we create a quantum circuit, denoted as $U(\theta)$, which is determined by some rule or parameterized by certain parameters $\theta$ that depend on the input $x^{(i)}$. This circuit is used to encode information from the input data into a quantum state. We use the quantum circuit $U(\theta)$ to prepare an input quantum state, denoted as $|\psi(x^{(i)})\rangle$, which carries the information embedded from the input data $x^{(i)}$. A multiply gate, denoted as $M(\theta)$, which depends on the parameter $\theta$, is applied to the input state $|\psi(x^{(i)})\rangle$ to obtain the output state $|\phi(x^{(i)})\rangle$. The measurement step involves measuring some observable under the output state $|\phi(x^{(i)})\rangle$. For example, we might measure the expectation value of the first qubit, denoted as $\langle \hat{O}_1 \rangle$. We define a function $F$, which can be a sigmoid function, softmax function, or a constant function, etc. The output of the quantum model, denoted as $y_{\text{model}}^{(i)}$, is computed as $F(\langle \hat{O}_1 \rangle)$. To assess the performance of our model, we calculate the cost function, denoted as $J(\theta)$, representing the divergence between the correct data $y^{(i)}$ and the output of the model $y_{\text{model}}^{(i)}$. This helps us quantify the error in our predictions. To improve the model's performance, we optimize the parameters $\theta$ to minimize the cost function $J(\theta)$. This is typically done using optimization algorithms like gradient descent or other suitable methods. Once the optimization process converges, we obtain the quantum circuit with optimized parameters $\theta$, denoted as $U(\theta_{\text{opt}})$. This trained quantum circuit serves as our desired prediction model for quantum data.

As a benchmark model of QFTs, we work on the Schwinger model (1+1d QED)~\cite{Schwinger:1962tp}, whose action is
\begin{align}
 S = \int d^2x\left[-\frac{1}{4} F^{\mu\nu} F_{\mu\nu} + \frac{g\theta}{4\pi}\epsilon^{\mu\nu}F_{\mu\nu} + \bar{\psi}(i\slashed{D}-m)\psi\right],
\end{align}
with $\slashed{D}=\gamma^\mu(p_\mu-i gA_\mu)$.
Here, $A_\mu$ is the $U(1)$ gauge potential, $E=\dot{A}_1$ is the corresponding electric field, $\psi$ is a two-component fermion field, $m$ is the fermion mass and $\gamma^\mu$ are two-dimensional $\gamma$-matrices. 

The Hamiltonian in temporal gauge $A_0=0$ is
\begin{align} \label{eq:Ham_conti}
H=\int dz \Big[    
    \frac{E^2}{2}
    -\bar{\psi}(i\gamma^1\partial_1 - g\gamma^1A_1 - me^{i\gamma_5\theta})\psi \Big] \,,
\end{align}
where the space-time coordinate is labeled by $x^\mu=(t,z)$. The Hamiltonian in the qubit representation is 
\begin{align}
\begin{aligned}
\label{eq:Ham_0}
    H=&\frac{1}{4a}\sum_{n=1}^{N-1}(X_nX_{n+1}+Y_nY_{n+1})\\
    &+\frac{m}{2}\sum_{n=1}^N(-1)^nZ_n+\frac{ag^2}{2}\sum_{n=1}^{N-1}L^2_n,
\end{aligned}
\end{align}
where $L_n$ is the electric field operator 
\begin{equation}
    L_n=\sum_{k=1}^n\frac{Z_k+(-1)^k}{2}.
\end{equation}
When simulating kinetic terms and electromagnetic fields in quantum circuits, an enormous amount of control gates are used. It is important to note that $\sum_{n=1}^{N-1}L^2_n$ contains the fully connected term, which makes computation noisy and heavy. Naively simulating a $N$-qubit system requires $N$ qubits, however, one can reduce the number of qubits and even gate depth by using QCL as we will demonstrate below. See Appendix for a detailed description of the model and operator definitions.

\if{
As shown in Fig.~\ref{fig:matrix} (upper), the Hamiltonian is sparse, although the model is fully connected. Interestingly, this model becomes more sparse as the system size increases. The ratio $\frac{\#\{H_{ij}:H_{ij}\neq0\}}{\#\{H_{ij}\}}$ of non-zero matrix components out of all matrix components ($\#\{H_{ij}\}=2^N\times 2^N$) is shown in Fig~\ref{fig:matrix} (lower left), which decays rapidly as $N$ increases. The reason for this is that the kinetic term that composes the off-diagonal components of the Hamiltonian is the nearest-neighbor interaction. Therefore most of the off-diagonal terms are zero. In fact, the number of non-zero off-diagonal components grows linear in the dimension of the Hilbert space ($2^N$), as shown in Fig.~\ref{fig:matrix} (lower right).
\begin{figure}
    \centering
    \includegraphics[width=0.49\linewidth]{Schwinger_Hamiltonian_matrix_N4_M1.png}
    \centering
    \includegraphics[width=0.49\linewidth]{Schwinger_Hamiltonian_matrix_N6_M1.png}
    \centering
    \includegraphics[width=0.49\linewidth]{Schwinger_M1_NonZeroComponents.pdf}
    \centering
    \includegraphics[width=0.49\linewidth]{Schwinger_M1_OffDiagonal_NonZeroComponents.pdf}
    \caption{Upper: Matrix components of the Scwiger model ($N=4,6$). Lower left: $N$-dependence of the ratio of non-zero matrix components out of all matrix components. Lower right: $N$-dependence of the number of non-zero off-diagonal components}
    \label{fig:matrix}
\end{figure}
}\fi

\emph{Benchmark Results.}---
As a simple demonstration, we study the real-time dynamics
\begin{equation}
    \ket{\psi(t)}=\mathcal{T}e^{-i\int_0^tdt'H}\ket{\psi_0},
\end{equation}
where $\ket{\psi_0}$ is the N\'{e}el state $\ket{1010\cdots10}$, which is the ground state of the Hamiltonian~\eqref{eq:Ham_0} at the large mass limit. For a given observable $0$, our training data set is a set of measurement results $O(t)\equiv\langle\psi(t)|O|\psi(t)\rangle$ at randomly chosen times $t$ and our objective is to learn the target Hamiltonian by comparing $O(t)$. The time-evolution of the chiral condensate density $\frac{1}{N}\langle\bar{\psi}\psi(t)\rangle$ is shown in Fig.~\ref{fig:Neel} (left). This data was obtained by the exact classical method for the $N=18$ system. Some data $\{x_i\}$ were sampled and used for QCL teaching data. The data was encoded into an initial quantum state $\ket{\varphi(x_i)}$ of QCL, where the state was updated by applying a unitary operator $U(\theta)$ to $\ket{\varphi(x_i)}$. Then the parameter $\theta$ is updated by optimizing the cost function. It should be emphasized that a three-qubit circuit was capable of predicting the dynamics in the $N=18$ system. The initial prediction of the chiral condensate is shown by the dashed line in Fig.~\ref{fig:Neel} (left) and the final prediction date is shown in the solid line. Moreover we also performed the same task for predicting the real-time dyamanimcs of the electric field. The teacher data, initial prediction result and final prediction result are shown in Fig.~\ref{fig:Neel} (left). Note that the electric field is a fully connected operator, therefore a precise measurement of the electric field operator is expensive in general, due to significant noise. Accurate measurements of electric field operators by ordinary quantum simulations are susceptible to noise. In contrast, QCL requires only one quantum operator to be measured and can be performed by small quantum circuits, giving it a practical advantage over the conventional quantum simulation of QFT.

\begin{figure}
    \centering
    \includegraphics[width=\linewidth]{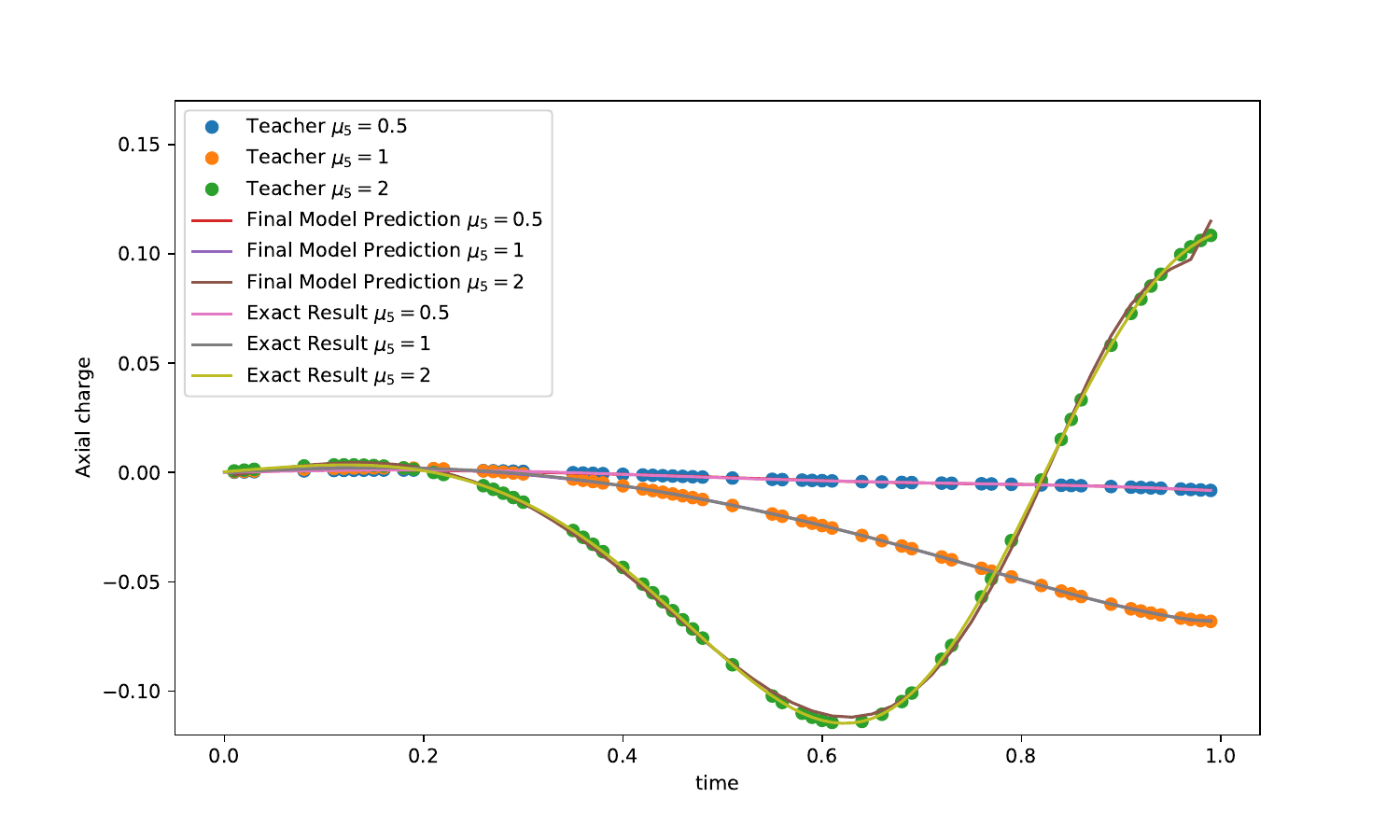}
    \caption{Prediction of the real-time dynamics of the axial charge after the quench.}
    \label{fig:axial}
\end{figure}

\emph{Predicting the chiral dynamics.}---
To perform a more complex task, we perform prediction of the real-time chiral dynamics. Again, we work on the massive Schwinger model with a finite chiral potential $\mu_5$~\cite{Kharzeev:2020kgc}. We prepare the initial state as the ground state of the following Hamiltonian 
\begin{align} \label{eq:Ham2}
H=\int dz \Bigg[    
    \frac{E^2}{2}
    -\bar{\psi}\left(i\gamma^1\partial_1 - g\gamma^1A_1-\gamma^1\frac{\dot{\theta}}{2} - me^{i\gamma_5\theta}\right)\psi \Bigg]
\end{align}
where $\theta=-2\mu_5t_0$. The difference from the previous model~\eqref{eq:Ham_conti} is the presence of the term $\mu_5\bar{\psi}\gamma_1\psi(=\mu_5Q_5)$, which induces the chiral imbalance in the initial state. This model is useful for discussing macroscopic quantum phenomena caused by the chiral anomaly, namely chiral magnetic effects and chiral magnetic waves. We evolve the state by the Hamiltonian $H(\mu_5=0,\theta=0)$, so the real-time evolution of the initial state is given by the time-ordered integral 
\begin{equation}
\label{eq:real_time_evolution}
    \ket{\psi(t)}=\mathcal{T}[e^{-i\int_0^tdt'H(\mu_5=0,\theta=0)}]\ket{\psi(0)}.
\end{equation}
The real-time evolution of the axial charge density (equivalently the vector current density) $\frac{1}{N}\langle Q_5(t)\rangle$ is shown in Fig.~\ref{fig:axial}. The three plots are labels by $\mu_5=0.5,1,2$, respectively. As before, the teacher data were obtained from rigorous simulations with $N=18$ systems, and prediction was performed with a five-qubit circuit. The use of large masses induces nonlinear rapid oscillations of the axial charge~\cite{Ikeda:2023vfk}. 

\begin{figure*}
    \centering
    \includegraphics[width=0.49\linewidth]{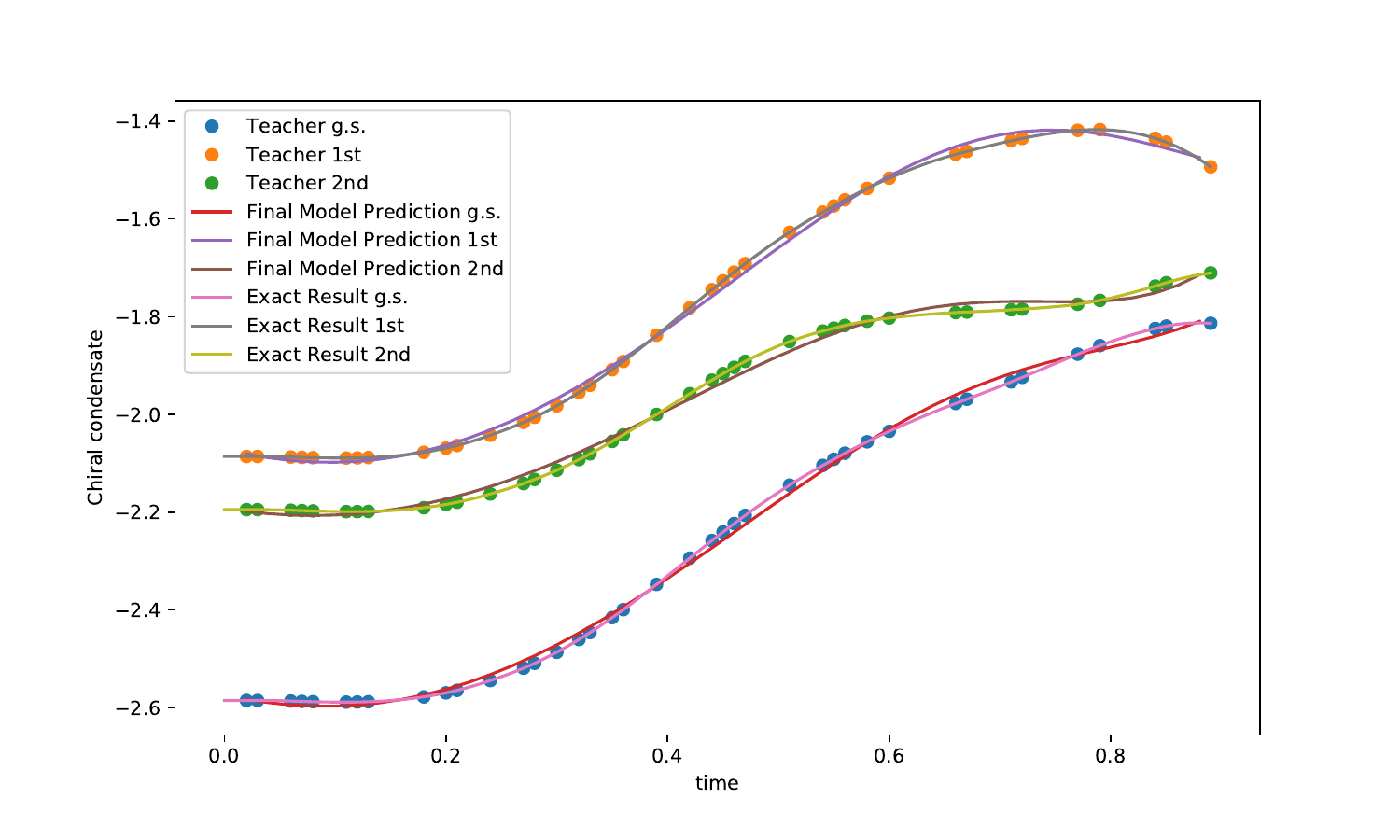}
    \centering
    \includegraphics[width=0.49\linewidth]{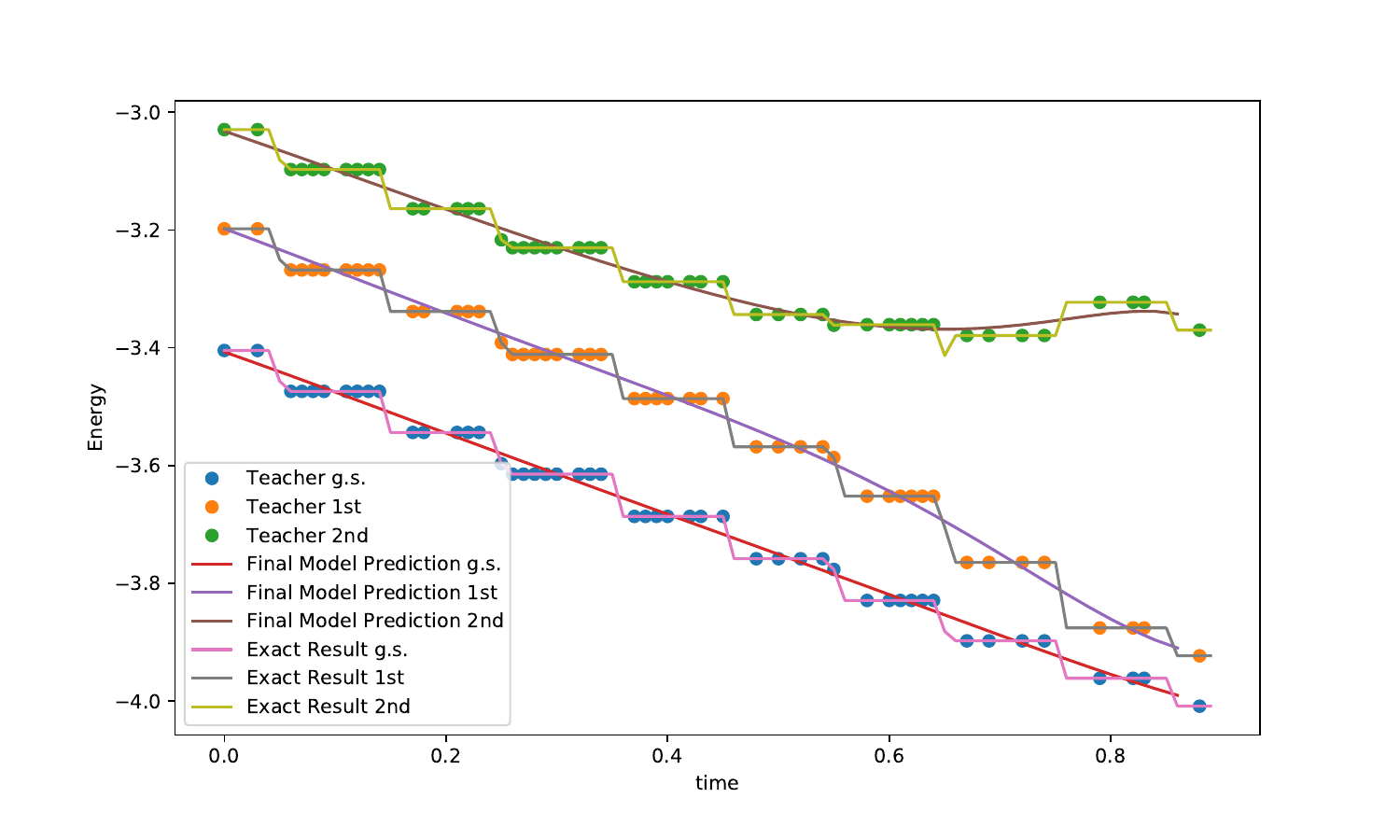}
    \caption{Prediction of the real-time evolution of the chiral condensate [left] and energy [right]. g.s., 1st and 2nd mean that the ground, first and second excited states are used as initial states.}
    \label{fig:jet}
\end{figure*}

\emph{Jet dynamics.}---
Here, using the massive Schwinger model coupled to external sources, we predict the quantum simulation of jet production using QCL. This is an extremely non-trivial task because of non-perturbative effects.  

We use the massive Schwinger model Hamiltonian~\eqref{eq:Ham_conti} in the presence of an external current $\jext^{\mu}$ describing the produced jets~\cite{Casher:1974vf,PhysRevLett.131.021902}:
\begin{align}
\begin{aligned}
H &= H_0 + H_1 ,\label{eq:Ham}\\
H_0 &=\int dx \left [ \frac{1}{2}E^2+\bar{\psi}(-i\gamma^1\partial_1+g\gamma^1A_1+m)\psi \right ], \\
H_1 &= \int dx\, \jext^1 A_1.
\end{aligned}
\end{align}

The effect on the theory of the interaction with the external source $H_1$ is to modify Gauss law to  
\begin{equation}
\label{eq:gauss}
\partial_1 E - j^0 = \jext^0 \ .
\end{equation}
with $j^0 = g\,\bar\psi\gamma^0\psi$.

To simulate the creation of a pair of jets in the context of $e^+e^-$ annihilation, we opt for an external current that represents charges of opposite polarity moving apart along the light cone. This external current can be defined as follows:

\begin{align}
\begin{aligned}
\jext^0(x,t)=&g[\delta(\Delta x - \Delta t) - \delta(\Delta x + \Delta t)]\Theta(\Delta t), \label{eq:jext}\\
\jext^1(x,t)=&g[\delta(\Delta x - \Delta t) + \delta(\Delta x + \Delta t)]\Theta(\Delta t) \ ,
\end{aligned}
\end{align}
where ($t_0$, $x_0$) represents the time and position of the point where the jet pair is generated, while $\Delta x \equiv x - x_0$ and $\Delta t \equiv t - t_0$ denote the spatial and temporal separation from this location. $\Theta$ is the Heaviside step function.
\begin{figure}
    \centering
    \includegraphics[width=\linewidth]{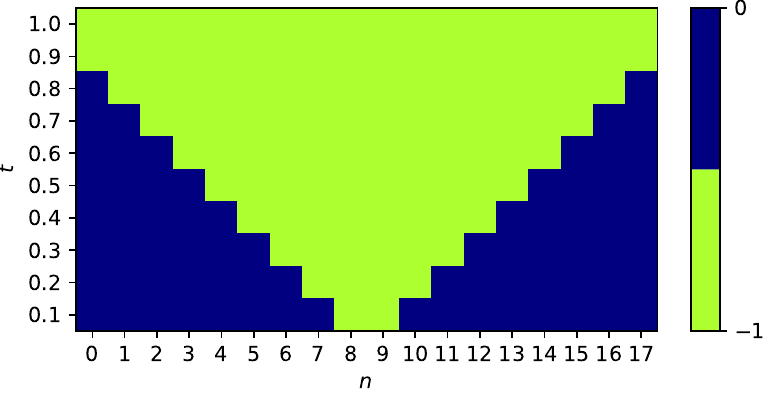}
    \caption{Behavior of $\Lext{n}(t)$, where a jet pair is produced at $t_0=0,x_0=N/2$.}
    \label{fig:Lext}
\end{figure}

The electric field is time-dependent and can be rewritten as $L_n=\Ldyn{n} + \Lext{n}$ and the Gauss law~\eqref{eq:gauss} is solved as follows:
\begin{align}
\begin{aligned}
  \Ldyn{n} &= \sum_{i=1}^n Q_i\,, \\
  \Lext{n}(t) &= -\Theta\left(t-a|n-N/2+1/2|)\right).
\end{aligned}
\end{align}
The time-dependent term of the electric field is shown in Fig.~\ref{fig:Lext}, which induces the propagation of the chiral condensate on the light cone.
The non-locality is contained in the dynamical gauge field and the external sources create a chain of electric fluxes between them.

The Hamiltonian is 
\begin{align}
     H(t)=&\frac{1}{4a}\sum_{n=1}^{N-1}(X_n X_{n+1}+Y_n Y_{n+1})+\frac{m}{2}\sum_{n=1}^N (-1)^n Z_n \notag\\
     &
     +\frac{a g^2}{2}\sum_{n=1}^{N-1} (\Ldyn{n}+\Lext{n}(t))^2 \ .
 \end{align}

Our simulations proceed as follows. We start by finding the eigenstate $\ket{\Psi_0}$ of the usual massive Schwinger model $H(t=0)$. We used the ground state, the 1st excited state or the 2nd excited state for $\ket{\psi_0}$. We then compute the state $\ket{\psi_t} = \mathcal{T} e^{-i \int_0^t H(t') \mathrm{d}t'}\ket{\psi_0}$ corresponding to the evolution under the time-dependent Hamiltonian $H(t)$. The system is effectively ``quenched" at $\frac{t}{a}=\frac{t_0}{a}=1$, when the external sources are introduced. 

To reproduce the results in~\cite{PhysRevLett.131.021902}, the data of $a=0.1,N=18,g=0.5/a,m=0.25/a$ obtained by the exact classical method were used as teaching data and predictions were made with a five-qubit QCL. The results are shown in Fig.~\ref{fig:jet}, where the left panel shows the time-evolution of the chiral condensate density $\langle\bar{\psi}\psi(t)\rangle$ whose initial states are ground state, the 1st excited state and 2nd excited state, and the right panel shows the energy expectation values $\frac{1}{N}\langle H(t)\rangle$. For both chiral condensation and energy prediction, only one qubit of measurement is required in QCL.

\emph{Conclusion.}---
In this research, quantum circuit learning is utilized to simulate quantum field theories (QFTs). Traditional quantum computer simulations of QFTs face challenges due to technical limitations, especially when implementing the Hamiltonian with Pauli spin matrices. To overcome these challenges, the study employs a small qubit configuration and low-depth quantum circuits to predict real-time dynamics in QFTs. One notable advantage is the ability to use a single-qubit measurement to accurately predict various physical parameters. The method is demonstrated to be effective by predicting quench dynamics, chiral dynamics, and jet production in a 1+1-dimensional model of quantum electrodynamics, closely matching results from classical calculations. This approach showcases the potential for efficiently simulating large-scale QFTs using near-term quantum devices, combining quantum and classical techniques.

\section*{Acknowledgment}
The author thanks Dmitri Kharzeev, 
C.R. Ramakrishnan, Shuzhe Shi and Ranjani Sundaram for useful discussion. This work was supported by the U.S. Department of Energy, Office of Science, National Quantum Information Science Research Centers, Co-design Center for Quantum Advantage (C2QA) under Contract No.DE-SC0012704.

\bibliography{main}
\appendix
\clearpage
\begin{widetext}

\section{The lattice Hamiltonian of the massive Schwinger model}
The Lagrangian density of the Schwinger model~\cite{Schwinger:1962tp} is 
\begin{equation}
\label{eq:L0}
    \mathcal{L} = -\frac{1}{4}F_{\mu\nu}F^{\mu\nu}+\bar{\psi}(i\gamma^\mu\partial_\mu-g\gamma^\mu A_\mu-m)\psi. 
\end{equation}
Here, we represent spacetime coordinates as $x^\mu=(t,z)$ and employ the following Dirac matrix notation: $\gamma^0 = Z$, $\gamma^1 = i,Y$, and $\gamma^5=\gamma^0 \gamma^1 = X$. In the context of $(1+1)$ dimensions, we establish the relationship between the axial charge density $Q_5(x)\equiv\bar{\psi}\gamma^5\gamma^0\psi(x)$ and the vector current density $J(x)\equiv\bar{\psi}\gamma^1\psi(x)$ as $Q_5(x) = -J(x)$. Similarly, the vector charge density $Q(x)\equiv\bar{\psi}\gamma^0\psi(x)$ and the axial current density $J_5(x)\equiv\bar{\psi}\gamma^5\gamma^1\psi(x)$ are linked by $Q(x) = J_5(x)$.

To discretize our Hamiltonian, we use staggered fermions~\cite{Kogut:1974ag, Susskind:1976jm}
\begin{equation}
    \psi_1(x)\to\frac{\chi_{2n}}{\sqrt{a}},~\psi_2(x)\to\frac{\chi_{2n+1}}{\sqrt{a}},
\end{equation}
where $a$ represents the finite lattice spacing. The lattice Hamiltonian corresponding to eq.~\eqref{eq:L0} is expressed as
\begin{align}
\begin{aligned}
\label{eq:lattice_total_ham3}
H=&-\frac{i}{2a}\sum_{n=1}^{N-1}
\big[U_n^\dag\chi^\dag_{n}\chi_{n+1}-U_n\chi^\dag_{n+1}\chi_{n}\big]
\nonumber\\
&+\frac{ag^2}{2}\sum_{n=1}^{N-1}L_n^2+ m\sum_{n=1}^{N} (-1)^n \chi^\dag_n\chi_n,
\end{aligned}
\end{align}
$U_n$ denotes the gauge link operator, and $L_n$ is the electric field operator that satisfies Gauss' law constraint described by the following equation:
\begin{equation}
    L_{n}-L_{n-1} =  \chi_n^\dagger \chi_n-\frac{1-(-1)^n}{2}.
    \label{eq:gauss_staggered}
\end{equation}
The link operator is written as $U_n = e^{-iagA_1(an)}$, using a lattice vector potential $\phi_n=ag,A_1(an)$.

For the purpose of quantum simulation, we transform the lattice Hamiltonian into the spin representation through the Jordan--Wigner transformation~\cite{Jordan:1928wi}:
\begin{align}
\begin{aligned}
 \chi_n=\frac{X_n-iY_n}{2}\prod_{i=1}^{n-1}(-i Z_i).
\end{aligned}
\end{align}
This transformation results in the Hamiltonian of the model becoming:
\begin{align}
\begin{aligned}
     H=&\frac{1}{8a}\sum_{n=1}^{N}\Big[(U_n+U^\dagger_n)\otimes(X_n X_{n+1}+Y_n Y_{n+1})
     +i(U_n-U^\dagger_n)\otimes(X_nY_{n+1}-Y_nX_{n+1})\Big]\\
     &+\frac{m}{2}\sum_{n=1}^N(-1)^n Z_n+\frac{a\ g^2}{2}\sum_{n=1}^{N} L^2_n.
\end{aligned}
\end{align}
We can eliminate the gauge link $U_n$ through a gauge transformation~\cite{Ikeda:2020agk}.

The local vector and axial charge densities are represented as follows:
\begin{align}
    Q_n \equiv \,& \bar{\psi}\gamma^0\psi = \frac{Z_n+(-1)^n}{2a},\\
    Q_{5,n} \equiv \,& \bar{\psi}\gamma^5\gamma^0\psi = \frac{X_nY_{n+1}-Y_nX_{n+1}}{4a}\,.
\end{align}
We define the total charge operator $Q \equiv a\sum_{n=1}^{N} Q_n$, which commutes with the Hamiltonian.
Assuming the boundary condition $L_0=0$, the Gauss' law constraint~\eqref{eq:gauss_staggered} leads to the solution:
\begin{align}
L_n = a\sum_{j=1}^n Q_j,.
\end{align}

\end{widetext}
\end{document}